\documentclass[10pt,aps,prd,twocolumn,showpacs,amsmath,amssymb,nofootinbib,eqsecnum,preprintnumbers,longbibliography]{revtex4-1}

\usepackage[english]{babel}				
\usepackage[mathscr]{eucal} 			
\usepackage{accents}					
\usepackage{mathtools}					
\usepackage[ddmmyyyy]{datetime}    		
\usepackage{enumitem}					
\usepackage{xcolor}						

\newcommand\bs[1]{\boldsymbol{#1}}

\newcommand\dd{\mathrm{d}}
\newcommand\pp{\partial}
\newcommand\de{\delta}

\newcommand\eps{\epsilon}

\newcommand\pois[2]{\left\{ #1\,{,}\,#2 \right\}}

\newcommand\dir[2]{\left\{ #1\,{,}\,#2 \right\}^*}

\newcommand\feq{\mathrel{\phantom{=}}}

\newcommand{\bwedge}{\bs{\wedge}}

\newcommand{\bdot}{\bs{\cdot}}

\newcommand\undervec[1]{\underaccent{\vec}{#1}}

\DeclareMathOperator{\sgn}{sgn}

\begin{document}


\title{Hamiltonian for scalar field model of infinite derivative gravity}

\author{Ivan Kol\'a\v{r}}
\email{i.kolar@rug.nl}
\affiliation{Van Swinderen Institute,
University of Groningen,
9747 AG, Groningen, The Netherlands}

\author{Anupam Mazumdar}
\email{anupam.mazumdar@rug.nl}
\affiliation{Van Swinderen Institute,
University of Groningen,
9747 AG, Groningen, The Netherlands}


\date{\today}

\begin{abstract}
Theories with an infinite number of derivatives are described by non-local Lagrangians for which the standard Hamiltonian formalism cannot be applied. Hamiltonians of special types of non-local theories can be constructed by means of the ${(1{+}1)}$-dimensional Hamiltonian formalism. In this paper, we consider a simple scalar field model inspired by the infinite derivative gravity and study its reduced phase space by using this formalism. Assuming the expansion of the solutions in the coupling constant, we compute the perturbative Hamiltonian and the symplectic 2-form. We also discuss an example of a theory leading to an infinite-dimensional reduced phase space for a different choice of the form factor.
\end{abstract}

\maketitle


\section{Introduction}

A recurrent feature that appears in many theories of quantum gravity is the non-locality. For example, string theory \cite{Polchinski:1998rr} is inherently non-local even on the classical level because strings and branes cannot interact at a specific point, but rather over a certain region. Similarly, there exists a minimal area in loop quantum gravity \cite{Ashtekar:2012np},  and causal set approach~\cite{Henson:2006kf}, which is expected to give rise to non-local behavior. It is not surprising that the non-localities appear also in the effective descriptions of string field theory \cite{Witten:1985cc,Freund:1987kt,Brezin:1990rb,Gross:1989ni,deLacroix:2017lif} and $p$-adic string theory \cite{Brekke:1988dg,Frampton:1988kr,Ghoshal:2000dd,Sen:2004nf,Biswas:2004qu,Ghoshal:2006te} (giving rise to zeta strings \cite{Dragovich:2007wb}). In these models, the Lagrangians contain kinetic operators with infinite number of derivatives.

It was realized already in \cite{Yukawa:1950a,Yukawa:1950b,Efimov:1967pjn} that the presence of infinite derivatives in the action may improve the ultraviolet behavior of loop integrals in many quantum field theories. This happens if the form-factors with an infinite number derivatives appearing in the actions are entire functions with no additional zeros in the complex plane. Later, it was demonstrated in \cite{Moffat:1990jj,Evens:1990wf,Tomboulis:1997gg} that gauge theories and gravity theories can be made ghost-free.

In fact, in the context of gravity, there exist concrete criteria for which the theories have the same number of degrees of freedom as there are in Einstein's theory when perturbed around particular backgrounds \cite{Moffat:2010bh,Biswas:2011ar,Barvinsky:2012ts,Biswas:2013kla,Barvinsky:2014lja,Biswas:2016egy,Mazumdar:2018xjz,Frolov:2015bia,Frolov:2015bta}. It turns out that such form-factors with an infinite number derivatives not only improve the ultraviolet behavior of the theories \cite{Modesto:2011kw,Modesto:2014lga,Talaganis:2014ida} but also resolve the cosmological singularities \cite{Biswas:2005qr,Biswas:2010zk,Biswas:2012bp,Koshelev:2012qn,Koshelev:2018rau} as well as the black-hole singularities \cite{Biswas:2011ar,Edholm:2016hbt,Frolov:2015usa,Koshelev:2017bxd,Buoninfante:2018xiw,Koshelev:2018hpt,Buoninfante:2018rlq,Buoninfante:2018stt,Giacchini:2018wlf,Buoninfante:2018xif}.

The Hamiltonian description of the non-degenerate Lagrangian systems with a finite number of derivatives $n$ was found by Ostrogradsky already in \cite{Ostrogradsky:1850fid}. He realized that such systems have ${2n}$-dimensional phase space and gave a prescription for its canonical coordinates. More importantly, he showed that the Hamiltonians of such systems are unbounded if ${n>1}$, which explains why most fundamental equations in physics are of the second-order at most.

A possible way to evade Ostrogradsky's theorem is to consider non-local Lagrangians with an infinite number of derivatives. By introducing derivatives of an arbitrary order, one might naively expect that we would need to prescribe an infinite number of initial conditions, however, this is not always the case. It is possible to find and solve differential equations with an infinite number of derivatives for which the initial value problem is well-defined with a finite number of initial data \cite{Barnaby:2007ve,Barnaby:2008tc,Gorka:2012hs,Carlsson:2015fnc}. The reason is because the initial data are often subject to infinitely many relations. As we will discus below, such relations are characterized by the constraints (in the Hamiltonian description) with an additional continuous parameter, which is identified as an extra dimension. 

A promising approach to counting the number of initial conditions and degrees of freedom without solving the differential equations is by means of the diffusion equation method \cite{Calcagni:2007ef,Calcagni:2007ru,Calcagni:2010ab,Calcagni:2018lyd,Calcagni:2018gke}. The advantage of this method is that it can be applied even to some non-linear theories if the non-localities are captured by the exponential form-factors.

The initial value problem is, however, best formulated in the Hamiltonian formalism. Focusing on special types of non-localities, it was found that one can rewrite a non-local Lagrangian theory as a local-in-time field theory with one extra dimension. Using the standard Legendre transformation, it is possible to arrive at the Hamiltonian formalism (with constraints) for such a field theory which is fully equivalent to the original non-local Lagrangian system. This formalism is referred to as the ${(1{+}1)}$-dimensional Hamiltonian formalism and it was first proposed in \cite{Llosa:1994}. Later, it was further developed and applied to various cases in \cite{Gomis:2000gy,Gomis:2000sp,Gomis:2003xv,Gomis:2004en}.\footnote{An equivalent formalism for non-localities of finite extent was introduced in \cite{Woodard:2000bt}. See also \cite{Bering:2000hc}, for an approach using boundary Poisson brackets.}

In particular, the ${(1{+}1)}$-dimensional Hamiltonian formalism was rewritten using constraints and applied to spacetime non-commutative theories \cite{Gomis:2000gy}. It was employed in the construction of gauge generators for spacetime non-commutative gauge theories \cite{Gomis:2000sp}. Finally, it was shown that this formalism can be used in the analysis of the perturbative reduced phase space and computation of the reduced Hamiltonian \cite{Gomis:2003xv,Gomis:2004en}. Here, the procedure was demonstrated on the $p$-adic string theory and the string field theory.

Hamiltonian formulation of the infinite derivative gravity is a very challenging task.\footnote{There were some attempts to develop the Hamiltonian formalism for the full infinite derivative gravity in the literature, e.g., \cite{Talaganis:2017,Joshi:2019cyk}. However, these approaches rely on infinite-dimensional generalization of the Ostrogradsky's formulas for canonical variables. In the ${(1{+}1)}$-dimensional formalism, on the other hand, only the standard definitions of canonical variables are used. Furthermore, in \cite{Talaganis:2017} the number of degrees of freedom was deduced from an indeterminate expression that was not evaluated by taking a proper limit.} In this paper, we restrict ourselves to a very simple scalar field model of the infinite derivative gravity and analyze its reduced phase space. The theory is constructed by perturbing the full action around the Minkowski background and assuming the scaling symmetry of equations of motion. We rewrite the theory in the ${(1{+}1}$-dimensional Hamiltonian formalism with constraints. By solving the second-class constraints we analyze the reduced phase space of the free theory and discuss the illustrative example where the phase space becomes infinite-dimensional. Considering the expansion of the solutions in the coupling constant, we compute the reduced Hamiltonian for the theory with the interaction term.

The paper is organized as follows: In Section~\ref{sc:hfnl}, we review the ${(1{+}1)}$-dimensional Hamiltonian formalism for non-local theories and discuss its limitations. In Section~\ref{sc:sfm}, we introduce a scalar field model of infinite derivative gravity. In Section~\ref{sc:ham}, we compute the perturbative Hamiltonian for the theory with the interaction term, and discuss an example of a theory leading to an infinite-dimensional phase space. The paper is concluded with a brief summary of the results in Sec.~\ref{sc:sum}. Appendix~\ref{ap:seccl} contains a supplementary proof of the second-class character of the constraints.


\section{Hamiltonian formalism for non-local theories} \label{sc:hfnl}

We begin by reviewing the ${(1{+}1)}$-dimensional Hamiltonian formalism for non-local theories following the works of \cite{Llosa:1994,Gomis:2000gy}. In addition, we comment on the limitations of this method and explain what types of non-local expressions are allowed by the formalism. We describe the formalism for single-variable systems. However, the extensions to multiple variables or field theories (with additional spatial dimensions) are also possible and they require minimal changes in formulas. We also briefly summarize necessary definitions of Dirac's procedure for constrained system \cite{Dirac:1958a,Dirac:1958b} (see also \cite{Dirac:1964,Henneaux:1994}) and describe the methods of constructing the reduced phase space using the ${(1{+}1)}$-dimensional Hamiltonian formalism \cite{Gomis:2003xv,Gomis:2004en}.

\subsection{Non-local Lagrangians}\label{ssc:nll}

Consider a system with a single variable $q(t)$ described by an action
\begin{equation}\label{eq:action}
S[q] = \int_\mathbb{R}\!\! d t\, L
\end{equation}
with the Lagrangian $L$. In standard local theories, $L$ is a function of $q(t)$ and its finitely many derivatives at time~$t$,
\begin{equation}
L=L(q(t),\dot{q}(t),\ldots,q^{(k)}(t))\;.
\end{equation}
In non-local theories, on the other hand, $L$ involves the dependence on variable $q$ at different times as well. Here, we will focus on Lagrangians that can be regarded as functionals ${L=L[q](t)}$ of the whole history of $q$ taking the particular form
\begin{equation}\label{eq:Lagr}
L = L[q(t+s)]\;.
\end{equation}
This means that $L$ is a $t$-dependent functional of a function $q$ that can only contain the expressions depending on $q(t+s)$, ${s\in\mathbb{R}}$.

This form includes the standard local terms with a finite number of derivatives as well as certain types of non-local terms such as the integrals over expressions with $q(t+s)$ or certain expressions with an infinite number of derivatives of $q(t)$. The reason is because differential operators $D(\pp_t)$ given by an analytic function $D(z)$ acting on $q(t)$ can be often rewritten as a convolution with an integral kernel $K(s)$,
\begin{equation}\label{eq:opF}
D(\pp_t)q(t) = \int_\mathbb{R}\!\! d s \, q(t+s) K(s)\;,
\end{equation}
which has the dependence of the form Eq.~\eqref{eq:Lagr}. Equality Eq.~\eqref{eq:opF} is generally expected to hold whenever the convolution exists, but it should be checked for each case separately, see for example \cite{Moeller:2002vx,Barnaby:2007ve}. It can be derived, for example, by repeated use of the Fourier transform,\footnote{Our convention for the Fourier transform is:
\begin{equation*}
\mathcal{F}[f](k) = \int_\mathbb{R}\!\!  d t\, f(t) e^{-i k t}\;,    
\quad
\mathcal{F}^{-1}[f](t) = \frac{1}{2\pi}\int_\mathbb{R}\!\! d k\, f(k) e^{i k t}\;.
\end{equation*}
} 
\begin{equation}
\begin{aligned}
D(\pp_t)q(t) &=\frac{1}{2\pi}\int_\mathbb{R}\!\!d k\, D(ik) \mathcal{F}[q](k) e^{ikt}
\\
&=\frac{1}{2\pi}\int_\mathbb{R}\!\!d k\, D(ik) \int_\mathbb{R}\!\!  d s\, q(s) e^{ik(t-s)}
\\
&=\frac{1}{2\pi}\int_\mathbb{R}\!\!d s\int_\mathbb{R}\!\!  d k\, D(-ik)e^{iks} q(t+s)\;,
\end{aligned}
\end{equation}
which gives an explicit formula for the integral kernel,
\begin{equation}
K(s) = \mathcal{F}^{-1}[D(-ik)](s)\;.
\end{equation}
Let us mention a few interesting examples of differential operators and their integral kernels,
\begin{equation}
\begin{aligned}
D(\pp_t) &= \pp_t^{l}\;,
& K(s) &= (-1)^l\de^{(l)}(s)\;,
\\
D(\pp_t) &= e^{a\pp_t}\;,
& K(s) &= \de(s-a)\;,
\\
D(\pp_t) &= e^{a\pp_t^2}\;,
& K(s) &= \tfrac{1}{2\sqrt{\pi a}}e^{-\frac{s^2}{4a}}\;,
\\
D(\pp_t) &= e^{a\pp_t^2}\pp_t^2\;,
& K(s) &= \tfrac{s^2-2 a}{8 \sqrt{\pi a^{5}} }e^{-\frac{s^2}{4 a}}\;,
\\
D(\pp_t) &= \sin(\pp_t^2)\;,
& K(s) &= \tfrac{1}{2 \sqrt{2 \pi }}\big[\sin(\tfrac{s^2}{4}){-}\cos(\tfrac{s^2}{4}) \big] \;.
\end{aligned}
\end{equation}
The first example is the $l^{\mathrm{th}}$ derivative $q^{(l)}(t)$. The second one is the shift operator in time corresponding to ${q(t+a)}$. The third operator appears in the effective models of $p$-adic string theory and the fourth one in string field theory. The last case is an example of $D(z)$ with infinite zeros in the complex plane. (This operator will be studied in Sec.~\ref{ssc:idof}.)

If the differential operator is not of the form $D(\pp_t)$, but contains the explicit temporal dependence, it seems unlikely that it could be expressed by means of ${q(t+s)}$. For instance, let us consider the operator ${e^{a t\pp_t}}$. When acting on $q(t)$, this operator scales the time by a constant $e^a$, ${e^{a t\pp_t}q(t)=q(e^a t)}$, instead of shifting it. Therefore, this non-locality is of a different type then what is assumed in Eq.~\eqref{eq:Lagr}. It may be possible to change the time coordinate and bring the expression to the form Eq.~\eqref{eq:opF}. For example, $e^{a t\pp_t}q(t)$ is just the time shift $e^{a \pp_{\tilde{t}}}\tilde{q}(\tilde{t})$ of the redefined function ${\tilde{q}(\tilde{t})=q(e^{\tilde{t}})}$ in the re-scaled coordinate ${\tilde{t}=\log{t}}$. However, such a transformation is beneficial only if it does not create unwanted non-localities in other parts of the Lagrangian.

Let us focus on the equations of motion for $q$. Using the chain rule for functional derivatives on the composed functional Eq.~\eqref{eq:action} with Eq.~\eqref{eq:Lagr}, the Euler--Lagrange equations can be rewritten in the form\footnote{We use the following notation for the functional differentiation: 
\begin{equation*}
	\de F[f;\de f]=\frac{d F[f+\eps \de f]}{d\eps}\Big|_{\eps=0}=\int_\mathbb{R}\!\!d x\,\frac{\de F[f]}{\de f(x)} \de f(x)\;,
\end{equation*}
where $\de F[f;g]$ denotes the variation and $\frac{\de F[f]}{\de f(x)}$ is the functional derivative.}
\begin{equation}\label{eq:ELeq}
\begin{aligned}
0 &= \frac{\de S}{\de q(t)} = \int_\mathbb{R}\!\! d s\, \frac{\de L(s)}{\de q(t)}\;.
\end{aligned}
\end{equation}
Due to the presence of non-local expressions possibly containing infinite number of derivatives, it is problematic to interpret the Euler--Lagrange equations as standard evolutionary equations for $q$ from initial data. They should be thought of as the functional relations constraining the whole function $q$ instead. Denoting the space of all possible trajectories $q(t)$ by $\mathcal{J}$, Eq.~\eqref{eq:ELeq} define the subspace of the physical trajectories ${\mathcal{J}_\mathrm{phys}\subset\mathcal{J}}$.

The equations of motion Eq.~\eqref{eq:ELeq} are usually very difficult to solve. However, if they can be recast into linear equations of the form
\begin{equation}\label{eq:linnonloc}
D(\pp_t)q(t) = j(t)\;,
\end{equation}
then the problem can be solved completely, see Refs.~\cite{Barnaby:2007ve,Barnaby:2008tc,Gorka:2012hs,Carlsson:2015fnc}. As shown in \cite{Barnaby:2007ve}, the full solution of \eqref{eq:linnonloc} can be found by means of the Laplace transform and the Cauchy integral theorem. The number of independent solutions is completely determined by the pole structure of $D(z)^{-1}$. This is because the operator $D(\pp_t)$ in the Laplace space reads\footnote{The Laplace transform is defined by formulas:
\begin{equation*}
\mathcal{L}[f](z) = \int_0^\infty\!\!  d t\, f(t) e^{-z t}\;,    
\quad
\mathcal{L}^{-1}[f](t) = \frac{1}{2\pi i}\!\oint_C \!\! f(z)e^{z t} d z\;,
\end{equation*}
where ${t\geq0}$ and $C$ encloses all the poles of the integrand.} 
\begin{equation}
D(\pp_t)q(t) = \frac{1}{2\pi i}\!\oint_C \!\! D(z)\mathcal{L}[q](z)e^{z t} d z\;,
\end{equation}
where the terms $q^{(l)}(0)$ are dropped as they can be absorbed into the arbitrary integration constants. The number of solutions is given by the homogeneous equation ${D(\pp_t)q(t)=0}$ which translates to the analyticity of $D(z)\mathcal{L}[q](z)$ due to the Cauchy integral theorem. This can be only satisfied by $\mathcal{L}[q](z)$ that has at most the same number of poles as $D(z)^{-1}$ of given or lower multiplicities. Each pole then contributes by a number of constants (independent solutions) that is equal to its multiplicity.


\subsection{${(1{+}1)}$-dimensional Hamiltonian formalism}

Let us consider the field quantities ${Q}$ that depend on an additional coordinate ${s\in\mathbb{R}}$, ${Q = Q(t,s)}$. We will denote the derivatives with respect to $t$ and $s$ by ${\dot{F}=\pp_t F}$ and ${F'=\pp_s F}$. The \textit{${(1{+}1)}$-dimensional Hamiltonian formalism} uses the definition of the Hamiltonian
\begin{equation}\label{eq:H}
H[Q,P](t) =\int_{\mathbb{R}}\!\!\dd s\,P(t,s)Q'(t,s)-\hat{L}[Q](t)\;,
\end{equation}
where $P(t,s)$ is the canonical momenta of $Q(t,s)$. The functional $\hat{L}$ is obtained from the Lagrangian $L$ by replacing ${q(t+s) \to Q(t,s)}$,
\begin{equation}
\hat{L}[Q](t) = L[q](t)\big|_{q(t+s)=Q(t,s)}\;, \quad s\in\mathbb{R}\;.
\end{equation}
This substitution effectively changes all non-local terms into the corresponding terms that are local in time $t$. Therefore, the Hamiltonian Eq.~\eqref{eq:H} defines an ordinary field theory that is local in time $t$. For instance, the non-local expressions of the form $D(\pp_t)q(t)$ are replaced by local-in-time terms (cf. Eq.~\eqref{eq:opF})
\begin{equation}\label{eq:DQ}
D(\pp_t)q(t)\big|_{q(t{+}s)=Q(t,s)} {=}\! \int_\mathbb{R}\!\! d s \, Q(t,s) K(s) 
{=} D(\pp_s)Q(t,0)\;.
\end{equation}

The phase space is defined as the cotangent bundle of all possible trajectories ${\mathcal{S}=\mathrm{T}^*\mathcal{J}}$ with the symplectic 2-form given by
\begin{equation}\label{eq:omega}
\Omega=\int_\mathbb{R} \!\! ds\,\dd Q(t,s)\bwedge \dd P(t,s)\;.
\end{equation}
The corresponding Poisson bracket of two phase-space observables ${F=F[Q,P]}$ and ${G=G[Q,P]}$ is
\begin{equation}\label{eq:pois}
\begin{aligned}
\pois{F}{G} &=\dd F \bdot \Omega^{-1} \bdot\dd G
\\
&= \int_{\mathbb{R}}\!\!d s\, \bigg[\frac{\de F}{\de Q(t,s)}\frac{\de G}{\de P(t,s)}-\frac{\de F}{\de  P(t,s)}\frac{\de G}{\de Q(t,s)}\bigg]\;.
\end{aligned}
\end{equation}
Hamilton's equations for \eqref{eq:H} are given by
\begin{equation} \label{eq:HE}
\begin{aligned}
\dot{Q}(t,s) &= \frac{\de H(t)}{\de P(t,s)} = Q'(t,s)\;,
\\
\dot{P}(t,s) &= -\frac{\de H(t)}{\de Q(t,s)} = P'(t,s) + \frac{\de \hat{L}(t)}{\de Q(t,s)}\;,
\end{aligned}
\end{equation}
where the first equation implies that ${Q(t,s)}$ is a function of the sum ${t+s}$, which is identified with ${q(t+s)}$. The second equation does not lead to further restrictions on $Q(t,s)$, meaning that the theory described by $H[Q,P]$ is not fully equivalent to the original theory associated with~$L[q]$.

As shown in Refs.~\cite{Llosa:1994,Gomis:2000gy}, the equivalent theory is obtained by restricting to the subspace ${\mathcal{S}_\mathrm{c}\subset\mathcal{S}}$ defined by the primary constraint,\footnote{As it is standard in Dirac's procedure for constrained systems, we use the \textit{weak equality} denoted by $\approx$ to emphasizes that the equation holds only on the constrained surface. The usual equality (with the standard sign $=$) is called the \textit{strong equality} and means that the equation holds everywhere in the phase space.}
\begin{equation}\label{eq:momc}
\Phi(t,s) =P(t,s) -\int_\mathbb{R}\!\!d \tilde{s}\, \chi(s,-\tilde{s})\frac{\de \hat{\mathscr{L}}(t,\tilde{s})}{\de Q (t,s)} \approx 0\;,
\end{equation}
where ${\chi(s,r){=}\tfrac12(\sgn(s){+}\sgn(r))}$ and $\hat{\mathscr{L}}(t,s)$ is a density-type functional
\begin{equation}
\hat{\mathscr{L}}[Q](t,s)=L[q](t)\big|_{q(t+\tilde{s})=Q(t,\tilde{s}+s)}\;, \quad \tilde{s}\in \mathbb{R}\;.
\end{equation}
Similar to \eqref{eq:DQ}, the non-local operators in this expression are replaced by local-in-time operators,
\begin{equation}\label{eq:simplerule}
D(\pp_t)q(t)\big|_{q(t{+}\tilde{s})=Q(t,\tilde{s}{+}s)} {=} D(\pp_s)Q(t,s)\;.
\end{equation}
Note that ${\hat{L}[Q](t)=\hat{\mathscr{L}}[Q](t,0)}$, which appears in the prescription for the Hamiltonian Eq.~\eqref{eq:H}. The constraint Eq.~\eqref{eq:momc} is called the \textit{momentum constraint} because it is associated with the definition of $P(t,s)$.

The consistency condition of the momentum constraint with the Hamiltonian evolution ${\dot\Phi\approx 0}$ generates the secondary constraint \cite{Gomis:2000gy},
\begin{equation}\label{eq:ELc}
\Psi(t,s) =\int_\mathbb{R}\!\!d \tilde{s}\, \frac{\de \hat{\mathscr{L}}(t,\tilde{s})}{\de Q (t,s)} \approx 0\;,
\end{equation}
which is referred to as the \textit{Euler--Lagrange constraint} because it corresponds to the Euler--Lagrange equation Eq.~\eqref{eq:ELeq} when ${Q(t,s)}$ is replaced by ${q(t{+}s)}$. This also establishes the equivalence between the dynamics of the Hamiltonian system with $H[Q,P]$ constrained on the surface $\mathcal{S}_\mathrm{c}$ and the space of physical trajectories $\mathcal{J}_\mathrm{phys}$ defined by the original non-local Lagrangian $L[q]$. When applied to the local theories with a finite number of derivatives, this formalism reproduces the Ostrogradsky's construction \cite{Ostrogradsky:1850fid}, see \cite{Llosa:1994,Gomis:2000gy}.

\subsection{Reduced phase space}
Arbitrary set of constraints can be always split into two classes. The \textit{first-class constraints} are such that their Poisson brackets with all other constraints weakly vanish. These constraints generate the gauge transformations (ignoring the cases when the Dirac's conjecture does not hold), which indicates that there is more then one set of canonical variables corresponding to a given physical state. 

The constraints that are not of the first class are called the \textit{second-class constraints}.\footnote{Do not confuse with the classification of \textit{primary} and \textit{secondary} constraints, which refers to the manner in which the constraints are generated from consistency with the Hamiltonian evolution.} These constraints are usually treated by replacing the standard Poisson bracket with the Dirac bracket. 

In this paper, we focus on the examples for which all constraints (momentum and Euler--Lagrange for various~$s$) form a second-class set. This means that the a pair of canonical variables $Q(t,s)$ and $P(t,s)$ that satisfies the constraints uniquely determines only one physical state. In this case, the matrix composed of all Poisson brackets of constraints with continuous indices $s$ and $\tilde{s}$ (and suppressed $t$-dependence),
\begin{equation}\label{eq:matrixC}
C(s,\tilde{s})=\begin{bmatrix}
\pois{\Psi(s)}{\Psi(\tilde{s})}\! & \!\pois{\Psi(s)}{\Phi(\tilde{s})}
\\
\pois{\Phi(s)}{\Psi(\tilde{s})}\! & \!\pois{\Phi(s)}{\Phi(\tilde{s})}
\end{bmatrix},
\end{equation}
has a maximal rank, so it can be inverted to find $C^{-1}$ by means of
\begin{equation}
\int_\mathbb{R}\!\! d\tilde{s}\, C(s,\tilde{s})\bdot C^{-1}(\tilde{s},\tilde{\tilde{s}})=\begin{bmatrix}
1 & 0
\\
0 & 1
\end{bmatrix}\de(s-\tilde{\tilde{s}})\;.
\end{equation}
The \textit{Dirac bracket} is then defined by the relation
\begin{equation}
\begin{gathered}
\dir{F}{G}=\pois{F}{G}-\int_\mathbb{R}\!\! ds\!\!\int_\mathbb{R}\!\! d\tilde{s}\,\undervec{F}(s)\bdot C^{-1}(s,\tilde{s})\bdot \vec{G}(\tilde{s})\;,
\\
\undervec{F}(s) {=}\begin{bmatrix}
\pois{F}{\Psi(s)}\! & \!\pois{F}{\Phi(s)}
\end{bmatrix},
\;\;
\vec{G}(\tilde{s}) {=}\begin{bmatrix}
\pois{\Psi(\tilde{s})}{G}
\\
\pois{\Phi(\tilde{s})}{G}
\end{bmatrix}.
\end{gathered}
\end{equation}
Since all the equations of the theory can be reformulated in terms of Dirac brackets, the second-class constraints effectively become strong equations expressing relation between canonical variables. This is because when working with Dirac brackets, we can set the second-class constraints equal to zero before evaluating the bracket. By solving the second-class constraints, we can entirely eliminate the redundant variables and, thus, determine the \textit{reduced phase space}. The \textit{number of degrees of freedom} of the theory is then equal to the half of the number of dimensions of the reduced phase space.

There are two different approaches to solving the second-class constraints and analyzing the reduced phase space \cite{Gomis:2003xv,Gomis:2004en}:
\begin{enumerate}[label=(\arabic*)]
\item Solve the Euler--Lagrange constraint ${\Psi\approx0}$, determine the momenta from ${\Phi\approx0}$, and find the expression for the symplectic 2-form on the constrained surfaces.
\item Expand the components in the Taylor frame Eqs.~\eqref{eq:compTayl1} and \eqref{eq:compTayl2}, find appropriate pairings between constraints $\Psi^k$ and $\Phi_l$ that eliminate the corresponding canonical pairs ${(q^j,p_j)}$.
\end{enumerate}
In this paper, we will focus on the former approach. The examples of both methods can be found in \cite{Gomis:2003xv}.


\section{Scalar field model of infinite derivative gravity}\label{sc:sfm}

It was shown in \cite{Biswas:2011ar,Biswas:2016etb,Biswas:2016egy} that the most general four-dimensional parity-invariant, torsion-free gravity action that is quadratic in curvature can be written in a rather compact form
\begin{equation}\label{eq:SQ}
\begin{aligned}
S_\textrm{QG}[g_{\mu\nu}] &=\frac{1}{2\kappa^2}\int_M\!\!\sqrt{-\mathfrak{g}}\Big[\mathcal{R}+\frac12\big(\mathcal{R}\mathcal{F}_1(\Box)\mathcal{R}
\\
&\feq+\mathcal{R}_{\mu\nu}\mathcal{F}_2(\Box)\mathcal{R}^{\mu\nu}+\mathcal{R}_{\mu\nu\kappa\lambda}\mathcal{F}_3(\Box)\mathcal{R}^{\mu\nu\kappa\lambda}\big)\Big]\;,
\end{aligned}
\end{equation}
where ${\kappa=\sqrt{8\pi G}}$ and the \textit{form-factors} $\mathcal{F}_i(\Box)$ are differential operators given by arbitrary analytic functions of d'Alembertian ${\Box=\nabla_\mu\nabla^\mu}$.\footnote{We do not consider non-analytic operators such as $\Box^{-1}$ \cite{Deser:2007jk,Conroy:2014eja}, or $\log(\Box)$ \cite{Barvinsky:1985an,Barvinsky:1993en,Donoghue:2014yha}.} Such actions are non-local if at least one form-factor is a non-polynomial function. In order for the theory to be ghost-free on the Minkowski background, the form factors are must satisfy: $2{\cal F}_1(\Box)+{\cal F}_2(\Box)+2 {\cal F}_3(\Box)=0$, see Ref.~\cite{Biswas:2011ar}. To simplify our lives, and without loss of generality, we can set ${\mathcal{F}_3(\Box)=0}$.\footnote{We should stress that the term involving $\mathcal{F}_3$ can be always neglected if one is interested in the second order metric perturbations, however, it contributes to the third order expansion.} Furthermore, we assume that the combination~\cite{Biswas:2011ar}
\begin{equation}
a(-\Box)\equiv 1-\mathcal{F}_1(\Box)\Box=1+\frac12 \mathcal{F}_2(\Box)\Box
\end{equation}
is an arbitrary entire function with no zeros in the complex plane satisfying ${a(0)=1}$ that can be expanded to all orders as
\begin{equation}
a(z)=1+\sum_{k=1}^\infty a_k z^k\;, \quad a_k=\frac{a^{(k)}(0)}{k!}\;.
\end{equation}
This ensure that the theory has the same number of dynamical degrees of freedom as general relativity and it reproduces the Einstein--Hilbert action in the local limit ${a(z) \to 1}$. The higher derivatives are suppressed by the \textit{scale of non-locality} $M_\textrm{s}$ which is implicitly included in the coefficients ${a_k\propto 1/M_\textrm{s}^{2k}}$.\footnote{The current bound on $M_\textrm{s}$ is ${\geq 0.004\,\mathrm{eV}}$. The constraint arises from the deviation of Newton's $1/r$ potential in torsion-based experiments~\cite{Edholm:2016hbt}.} With these assumptions, we can rewrite the action as
\begin{equation}\label{eq:SIDG}
\begin{aligned}
S_\textrm{IDG}[g_{\mu\nu}]&=\frac{1}{2\kappa^2}\int_M\!\!\sqrt{-\mathfrak{g}}\Big[\mathcal{R}-\mathcal{G}_{\mu\nu}\frac{1-a(-\Box)}{\Box}\mathcal{R}^{\mu\nu}\Big]\;,
\end{aligned}
\end{equation}
which belongs to the class of theories of the \textit{infinite derivative gravity}. Here, ${{\mathcal{G}}_{\mu\nu}=\mathcal{R}^{\mu\nu}{-}\frac12\mathcal{R}g_{\mu\nu}}$ is the Einstein tensor.

By perturbing this action around the Minkowski background,
\begin{equation}
g_{\mu\nu}=\eta_{\mu\nu}+\kappa h_{\mu\nu}\;,
\end{equation}
we can find that this theory is ghost-free and the only propagating degree of freedom is the massless spin-2 graviton. Indeed, the only pole of the propagator in the Fourier space,
\begin{equation}
\Pi(k)=\frac{\Pi_\textrm{GR}(k)}{a(k^2)}\;,
\end{equation}
is the pole ${k^2=0}$ corresponding to the graviton propagator of general relativity~$\Pi_\textrm{GR}(k)$ because $a(k^2)$ has no zeros in the complex plane \cite{Biswas:2011ar}.

Obviously, the complicated action Eg.~\eqref{eq:SIDG} is still way beyond the scope of applicability of the Hamiltonian formalism discussed in Sec.~\ref{sc:hfnl}. Note that we are forced to work with perturbative analysis because the formalism is not covariant and one cannot transform the non-local terms in the action to the form Eq.~\eqref{eq:opF} for a general metric $g_{\mu\nu}$. These coordinates might exist only for very special geometries such as the ultra-static spacetimes where ${\Box=-\pp_t^2+\Delta}$, with $\Delta$ being the Laplacian of the spatial part of the metric. 

The perturbative expansion of the action around the Minkowski background contains many terms with a complicated tensorial structure. To keep the problem manageable we focus on the terms involving only the trace ${h^\mu_\mu}$ and construct a scalar field theory for ${\phi\equiv h^\mu_\mu}$ whose equations of motion enjoy the same scaling symmetry as Einstein's equations. Such scalar field models of infinite derivative gravity were studied in \cite{Talaganis:2014ida,Talaganis:2016ovm,Buoninfante:2018gce,Buoninfante:2019swn}. We follow \cite{Talaganis:2014ida} where the scalar field action was constructed by examining all possible terms of the third order metric perturbations of \eqref{eq:SIDG}. Keeping the trace terms only, it was found that the free part $S_\textrm{free}$ and the interaction part $S_\textrm{int}$ of the scalar field action ${S=S_\textrm{free}+S_\textrm{int}}$ should take the following form:
\begin{equation}\label{eq:actionfreeint}
\begin{aligned}
S_\textrm{free}[\phi] &=\frac12\int\!\! d^4 x\, \phi a(-\Box)\Box\phi\;,
\\
S_\textrm{int}[\phi] &=\kappa\int\!\! d^4 x\, \big(\alpha_1\phi \pp_\mu\phi\pp^\mu\phi +\alpha_2\phi\Box\phi a(-\Box)\phi
\\
&\feq+\alpha_3\phi\pp_\mu\phi a(-\Box)\pp^\mu\phi\big)\;,
\end{aligned}
\end{equation}
where ${\Box=\pp_\mu\pp^\mu}$.

Unfortunately, the particular values of constants $\alpha_i$ are technically quite difficult to find. An alternative method (proposed in \cite{Talaganis:2014ida}) is to specify the coefficients $\alpha_i$ by demanding the equation of motion to have the same scaling symmetry ${g_{\mu\nu}\to (1+\eps)g_{\mu\nu}}$ as Einstein's equations. This translates into the transformation of the scalar field
\begin{equation}\label{eq:symphi}
\phi \to \tilde{\phi}=(1+\eps)\phi + \eps \kappa^{-1}\;.
\end{equation}
Since the transformations that scale the action leave the equations of motion invariant, we look for the coefficients $\alpha_i$ satisfying
\begin{equation}
\de S[\phi;\delta_\textrm{s}\phi]\propto S[\phi]\;,
\end{equation}
where we denoted
\begin{equation}
\de_\textrm{s}\phi=(\tilde{\phi}-\phi)/\eps=\phi+\kappa^{-1}\;.
\end{equation}
It is useful to split the action Eq.~\eqref{eq:actionfreeint} into local and non-local parts ${S=S_\textrm{loc}+S_\textrm{n-l}}$ and demand the proportionality for both cases with the same constant. We obtain the following relations:
\begin{equation}
\alpha_1=\alpha_2=\alpha_3+\frac12\;.
\end{equation}

We should stress that only the local part $S_\textrm{loc}$ is specified uniquely by this method. The non-local part $S_\textrm{n-l}$ still depends on one parameter. For simplicity, we choose\footnote{The authors of \cite{Talaganis:2014ida,Talaganis:2016ovm,Buoninfante:2018gce,Buoninfante:2019swn} studied the theory corresponding to the choice ${\alpha_1=\alpha_2=-\alpha_3=1/4}$. In our computation we have found that we can simplify the interaction term by making a judicious choice of Eq.~\eqref{alphai}.}
\begin{equation}\label{alphai}
\alpha_1=0\;,\quad \alpha_2=0\;,\quad \alpha_3=-\frac12\;,
\end{equation}
which gives rise to the action
\begin{equation}\label{eq:actionphi}
S[\phi]=\frac12\int\!\! d^4 x\, \phi a(-\Box)\Box\phi+\frac{\kappa}{4}\int\!\! d^4 x\, \phi^2 a(-\Box)\Box\phi \;.
\end{equation}
The computations presented in the next section could be extended to other choices of $\alpha_i$. 

In what follows, we will focus on the spatially homogeneous fields ${\phi=\phi(t)\equiv q(t)}$. Ignoring the spatial derivatives and re-scaling the action by ${\int\!\!d^3 x}$, we arrive at the non-local Lagrangian for a single variable $q(t)$,
\begin{equation}\label{eq:lagrangian}
	L[q](t) = -\frac12 q(t) a(\pp_t^2)\ddot{q}(t)-\frac{\kappa}{4}q(t)^2 a(\pp_t^2)\ddot{q}(t) \;,
\end{equation}
for which we compute the Hamiltonian using the ${(1{+}1)}$-dimensional formalism.


\section{Computation of Hamiltonian}\label{sc:ham}
Employing the formalism of Sec.~\ref{sc:hfnl}, we compute the Hamiltonian for the scalar field model introduced in Sec.~\ref{sc:sfm}. To provide a simple starting point, we begin at the level of the free theory where the non-locality does not play a significant role. Then we move on to the theory with interactions and compute the perturbative Hamiltonian. Finally, we study an example of a free theory leading to an infinite-dimensional reduced phase space.

\subsection{Free theory}

We start with the free theory given by the first term in Eq.~\eqref{eq:lagrangian}. The time-localized density-type functional $\hat{\mathscr{L}}[Q](t,s)$ for the ${(1{+}1)}$-dimensional field ${Q(t,s)}$ reads
\begin{equation}
	\hat{\mathscr{L}}[Q](t,s) = -\frac12 Q(t,s) a(\pp_s^2)Q''(t,s)\;,
\end{equation}
where we used the rule Eq.~\eqref{eq:simplerule}.

For our calculations it is useful to have the functional derivative of $\hat{\mathscr{L}}(t,s)$,
\begin{equation}
\begin{aligned}
\frac{\de \hat{\mathscr{L}}(t,s)}{\de Q (t,\tilde{s})} &=-\frac12\delta(s-\tilde{s})a(\pp_s^2)Q''(t,s)
\\
&\feq-\frac12 Q(t,s)a(\pp_s^2)\delta''(s-\tilde{s})\;.
\end{aligned}
\end{equation}
By integrating this expression (see Eq.~\eqref{eq:ELc}) we obtain the Euler--Lagrange constraint,
\begin{equation}\label{eq:psifree}
\Psi(t,s)=-a(\pp_s^2)Q''(t,s)\approx0\;.
\end{equation}
Employing the identity
\begin{equation}\label{eq:amid}
\chi(s,-\tilde{s})\de^{(n)}(\tilde{s}-s)=\sum_{k=0}^{n-1}(-1)^k \de^{(k)}(s) \,\de^{(n-k-1)}(\tilde{s})\;,
\end{equation}
we can obtain the momentum constraint (see Eq.~\eqref{eq:momc}),
\begin{equation}\label{eq:phifree}
\Phi(t,s)=P(t,s)-\frac12\sum_{k,j=0}^\infty a_{k+j}Z^{k,j}[Q](t,s)\approx0\;,
\end{equation}
where
\begin{equation}\label{eq:Zfunc}
\begin{aligned}
Z^{k,j}[Q](t,s) &=Q^{(2k+1)}(t,0)\,\de^{(2j)}(s)
\\
&\feq+Q^{(2k)}(t,0)\,\de^{(2j+1)}(s)\;.
\end{aligned}
\end{equation}
At this point, we should check that the constraints $\Psi(t,s)$ and $\Phi(t,s)$ form a second-class set. This is done in Appx.~\ref{ap:seccl} by expanding the constraints and variables in the Taylor basis and computing the matrix Eq.~\eqref{eq:matrixC}.

The linear differential equation Eq.~\eqref{eq:psifree} can be solved exactly. As we discussed in Sec.~\ref{ssc:nll}, the number of independent solutions of such equations is determined by the pole structure of the function $1/(a(z^2)z^2)$, which is the same as $1/z^2$ because $a(z^2)$ has no zeros in the complex plane. This means that all solutions of the Euler--Lagrange constraint Eq.~\eqref{eq:psifree} are also the solutions of a simpler constraint,
\begin{equation}\label{eq:psifreesimpler}
\tilde{\Psi}(t,s)=Q''(t,s)\approx0\;,
\end{equation}
which is solved by the linear function
\begin{equation}\label{eq:Qred}
Q(t,s)=q_0(t)+q_1(t)s\;.
\end{equation}
Inserting this expression in the momentum constraints Eq.~\eqref{eq:phifree}, we find 
\begin{equation}\label{eq:Pred}
P(t,s)=\frac12 q_1(t)\sum_{j=0}^\infty a_j\de^{(2j)}(s)+\frac12 q_0(t)\sum_{j=0}^\infty a_j\de^{(2j+1)}(s)\;.
\end{equation}

Therefore, we have shown that the phase-space variables ${(Q,P)}$ are described solely in terms of ${q_0=q}$ and ${q_1=\dot{q}}$. These quantities can be used to parametrize the reduced phase space of the system. Employing the relations Eqs.~\eqref{eq:H}, \eqref{eq:omega}, \eqref{eq:Qred}, and \eqref{eq:Pred}, we obtain the reduced Hamiltonian and the symplectic 2-form,
\begin{equation}\label{eq:hamomfree}
H_\mathrm{red}=\frac12 \dot{q}^2\;,
\quad 
\Omega_\mathrm{red}=\dd q\bwedge\dd \dot{q}\;.
\end{equation}
We can see from the form of the symplectic 2-form that the reduced phase space is two-dimensional. Thus, the theory has one degree of freedom. This confirms the expected result obtained by the inspection of the propagator. The system is dynamically equivalent to the original Lagrangian without the non-local operator, ${L = -\frac12 q \ddot{q}}$.

Note that we could alternatively obtain the same results by using more general formulas
\begin{equation}\label{eq:HOalter}
\begin{aligned}
H_\mathrm{red} &=\frac12 \sum_{k,j=0}^\infty a_{k+j}\big(Q^{(2k{+}1)}_{s{=}0}Q^{(2j{+}1)}_{s{=}0}-Q^{(2k)}_{s{=}0}Q^{(2j{+}2)}_{s{=}0}\big)\;,
\\
\Omega_\mathrm{red} &=\sum_{k,j=0}^\infty a_{k+j}\dd Q^{(2k)}_{s{=}0}\bwedge\dd Q^{(2j{+}1)}_{s{=}0}\;,
\end{aligned}
\end{equation}
where ${Q^{(m)}_{s{=}0}\equiv Q^{(m)}(t,0)}$. These expressions hold true for arbitrary analytic function $a(z)$. They can be derived by inserting $P(t,s)$ from Eq.~\eqref{eq:phifree} in Eqs.~\eqref{eq:H}, \eqref{eq:omega}, and using Eq.~\eqref{eq:psifree}.


\subsection{Interaction term}

Let us consider the Lagrangian Eq.~\eqref{eq:lagrangian} with the cubic interaction term. Using the rule Eq.~\eqref{eq:simplerule}, we can write the functional $\hat{\mathscr{L}}[Q](t,s)$ of the ${(1{+}1)}$-dimensional Hamiltonian formalism as
\begin{equation}
\begin{aligned}
	\hat{\mathscr{L}}[Q](t,s) &= -\frac12 Q(t,s) a(\pp_s^2)Q''(t,s)
	\\
	&\feq-\frac{\kappa}{4}Q(t,s)^2 a(\pp_s^2)Q''(t,s) \;.
\end{aligned}
\end{equation}
Its functional derivative reads
\begin{equation}
\begin{aligned}
\frac{\de \hat{\mathscr{L}}(t,s)}{\de Q (t,\tilde{s})} &=-\frac12\delta(s-\tilde{s})a(\pp_s^2)Q''(t,s)
\\
&\feq-\frac12 Q(t,s)a(\pp_s^2)\delta''(s-\tilde{s})
\\
&\feq-\frac{\kappa}{2}\delta(s-\tilde{s})Q(t,s)a(\pp_s^2)Q''(t,s)
\\
&\feq-\frac{\kappa}{4}Q(t,s)^2 a(\pp_s^2)\delta''(s-\tilde{s})\;.
\end{aligned}
\end{equation}
This expression can be integrated to obtain the Euler--Lagrange constraints with the additional non-linear terms in $Q$,
\begin{equation}\label{eq:ELint}
\begin{aligned}
\Psi(t,s) &=-a(\pp_s^2)Q''(t,s) -\frac{\kappa}{2}Q(t,s)a(\pp_s^2)Q''(t,s)
\\
&\feq-\frac{\kappa}{4}a(\pp_s^2)\pp_s^2(Q(t,s)^2)\approx0\;.
\end{aligned}
\end{equation}
With the help of the identity Eq.~\eqref{eq:amid}, we find momentum constraint
\begin{equation}\label{eq:momint}
\begin{aligned}
\Phi(t,s) &=P(t,s)-\frac12\sum_{k,j=0}^\infty a_{k+j} Z^{k,j}[Q](t,s)
\\
&\feq-\frac{\kappa}{4}\sum_{k,j=0}^\infty a_{k+j} Z^{k,j}[Q^2](t,s)\approx0\;,
\end{aligned}
\end{equation}
where $Z^{k,j}$ is the functional defined in Eq.~\eqref{eq:Zfunc}. The interaction term does not change the second-class character of the constraints $\Psi(t,s)$ and $\Phi(t,s)$.

Due to the non-linear terms in Eq.~\eqref{eq:ELint} we must resign to the perturbative approach in the coupling constant $\kappa$ to solving the Euler--Lagrange constraint. However, let us first discuss the solution of the local limit ${a(\pp_s^2)\to 1}$ (i.e., ${a_k\to0}$, ${k>0}$). The differential equation Eq.~\eqref{eq:ELint} reduces to
\begin{equation}
Q'' +\kappa Q Q''+\frac{\kappa}{2}(Q')^2=0\;,
\end{equation}
which has an exact solution
\begin{equation}\label{eq:exactsolloc}
Q_\textrm{loc}(t,s)=\big(\kappa^{-1}+q_0(t)\big)^{\frac13} \Big(\kappa^{-1}+q_0(t)+\frac32 q_1(t) s\Big)^{\!\frac23}\!-\kappa^{-1}\;,
\end{equation}
with two arbitrary functions ${q_0(t)}$, ${q_1(t)}$. Since all derivatives of this function can be continuously extended to ${\kappa=0}$, it seems reasonable to restrict ourselves to the solutions of the non-local problem that also admit smooth expansion in the coupling constant $\kappa$.

Following Ref.~\cite{Llosa:1994}, we define the \textit{perturbative solution} of Eq.~\eqref{eq:ELint} as a power series
\begin{equation}\label{eq:Qps}
Q(t,s)=\sum_{k=0}^\infty \kappa^k Q_{k}(t,s)\;,
\end{equation}
such that the initial data are given by a set of arbitrary functions $q_0(t)$ and $q_1(t)$,
\begin{equation}\label{eq:initialcond}
Q_{k}(t,0)=\delta_k^0\, q_0(t)\;,
\quad
Q_{k}'(t,0)=\delta_k^0\, q_1(t)\;,
\end{equation}
and the coefficients $\Psi_k(t,s)$ of series
\begin{equation}
\Psi(t,s) =\sum_{k=0}^\infty\kappa^k\Psi_k(t,s)\;,
\end{equation}
all vanish for such functions $Q(t,s)$. Inserting Eq.~\eqref{eq:Qps} in Eq.~\eqref{eq:ELint} we can write the equations for $Q_k(t,s)$ as
\begin{equation}\label{eq:psikint}
\begin{aligned}
\Psi_k &=-a(\pp_s^2)Q_k''
-\frac12\sum_{m=0}^{k-1} Q_m a(\pp_s^2)Q''_{k-m-1}
\\
&\feq-\frac14\sum_{m=0}^{k-1} a(\pp_s^2)\pp_s^2\big(Q_m Q_{k-m-1}\big)\approx0\;.
\end{aligned}
\end{equation}

Recall that for $Q_0$ corresponding to $\Psi_0$, the general solution of Eq.~\eqref{eq:psifree} was a solution of a much simpler equation Eq.~\eqref{eq:psifreesimpler}. This can be generalized to the constraints $\Psi_k$ with ${k>0}$ as well because the differential equations Eq.~\eqref{eq:psikint} are of the form Eq.~\eqref{eq:psifree} with the non-vanishing right-hand side given by $Q_{j}$ with ${j<k}$ (see Eq.~\eqref{eq:linnonloc}). A general solution of the constraints Eq.~\eqref{eq:psikint} is equivalent to the solution of non-homogeneous equations,
\begin{equation}\label{eq:simpELcint}
\begin{aligned}
\tilde{\Psi}_k &=Q_k''
+\frac12\sum_{m=0}^{k-1} \frac{1}{a(\pp_s^2)}\big(Q_m a(\pp_s^2)Q''_{k-m-1}\big)
\\
&\feq+\frac14\sum_{m=0}^{k-1} \pp_s^2\big(Q_m Q_{k-m-1}\big)\approx0\;.
\end{aligned}
\end{equation}
Here, the operator $1/a(\pp_s^2)$ is given by the analytic function $1/a(z)$. Its coefficients can be generated from the expansion around ${z=0}$,
\begin{equation}
\frac{1}{a(z)}=\sum_{k=0}^\infty \frac{(1/a)^{(k)}(0)}{k!}z^k=1-a_1 z+(a_1^2{-}a_2)z^2+\ldots\;.
\end{equation}

Solving Eq.~\eqref{eq:simpELcint} iteratively and taking into account initial conditions Eq.~\eqref{eq:initialcond}, we arrive at the perturbative solution of the Euler--Lagrange constraint Eq.~\eqref{eq:ELint},
\begin{equation}\label{eq:solElint}
\begin{aligned}
Q(t,s) &=q_0(t)+q_1(t) s-\tfrac14\kappa q_1(t)^2 s^2
\\
&\feq+\tfrac12\kappa^2 q_1(t)^2 s^2\big(\tfrac12 q_0(t) +\tfrac13 q_1(t) s\big)
\\
&\feq+\tfrac14\kappa^3 q_1(t)^2 s^2\big(\tfrac94 a_1 q_1(t)^2 -q_0(t)^2
\\
&\feq-\tfrac43 q_0(t) q_1(t) s -\tfrac7{12}q_1(t)^2 s^2\big) +\mathcal{O}(\kappa^4)\;.
\end{aligned}
\end{equation}
This can be compared with the Taylor series of exact local solution Eq.~\eqref{eq:exactsolloc},
\begin{equation}
Q=Q_\textrm{loc}+\frac9{16} \kappa^3 a_1 q_1^4 s^2 +\mathcal{O}(\kappa^4)\;,
\end{equation}
where we explicitly see the first non-local contribution. By inserting Eq.~\eqref{eq:solElint} in the momentum constraints Eq.~\eqref{eq:momint}, we obtain
\begin{equation}\label{eq:solMint}
\begin{aligned}
P(t,s) &=\tfrac12 \big(1{+}\kappa q_0(t)\big)q_1(t)\sum\nolimits_{j} a_j\delta^{(2j)}(s)
\\
&\feq+\tfrac12 \big(1{+}\tfrac12 \kappa q_0(t)\big)q_0(t)\sum\nolimits_{j} a_j\delta^{(2j{+}1)}(s)
\\
&\feq-\tfrac14 \kappa^2\big(1{-}\kappa q_0(t)\big)q_1(t)^3\sum\nolimits_{j} a_{j{+}1}\delta^{(2j)}(s)
\\
&\feq+\tfrac14 \kappa\big(1{+}\tfrac94 \kappa^2 a_1 q_1(t)^2\big)q_1(t)^2\sum\nolimits_{j} a_{j{+}1}\delta^{(2j{+}1)}(s)
\\
&\feq+\tfrac58 \kappa^3 q_1(t)^4\sum\nolimits_{j} a_{j+2}\delta^{(2j{+}1)}(s)+\mathcal{O}(\kappa^4)\;,
\end{aligned}
\end{equation}
where we denoted ${\sum_{j}\equiv\sum_{j=0}^\infty}$.

By this procedure, we constructed the two-dimensional reduced phase space of perturbative solutions that is parametrized by variables ${q_0=q}$ and ${q_1=\dot{q}}$. The Hamiltonian and the symplectic 2-form can be computed by inserting Eqs.~\eqref{eq:solElint} and \eqref{eq:solMint} in Eqs.~\eqref{eq:H} and \eqref{eq:omega}, 
\begin{equation}\label{eq:hamom}
\begin{aligned}
H_\mathrm{red} &=\frac12\dot{q}^2 +\frac12 \kappa q \dot{q}^2+\frac{3}{8}\kappa^2 a_1 \dot{q}^4-\frac{3}{8}\kappa^3 a_1 q\dot{q}^4+\mathcal{O}(\kappa^4)\;,
\\
\Omega_\mathrm{red} &=\Big(1{+}\kappa q{+}\frac{3}{2}\kappa^2 a_1\dot{q}^2{-}\frac{3}{2}\kappa^3 a_1 q\dot{q}^2 +\mathcal{O}(\kappa^4)\Big)\,\dd q\bwedge\dd \dot{q}\;.
\end{aligned}
\end{equation}
The first term with the non-local contribution appears in the second order of the coupling constant $\kappa$ in both expressions. 

However, we should remind that we restricted ourselves to the lowest-order interaction term in the action. If we included higher-order interactions as well, the Hamiltonian and the symplectic 2-form might get additional contributions. In such situation, we should be more careful with the order of $\kappa$ and write just
\begin{equation}
\begin{aligned}
H_\mathrm{red} &=\frac12\dot{q}^2 +\frac12 \kappa q \dot{q}^2+\mathcal{O}(\kappa^2)\;,
\\
\Omega_\mathrm{red} &=\big(1{+}\kappa q+\mathcal{O}(\kappa^2)\big)\,\dd q\bwedge\dd \dot{q}\;,
\end{aligned}
\end{equation}
which is equivalent to the local case at this order.


\subsection{Theory with infinite degrees of freedom}\label{ssc:idof}
So far we have considered $a(z)$ to be an entire function with no zeros in the complex plane. We have shown that the corresponding free theory has one degree of freedom and the phase space is two-dimensional. If this assumption is not satisfied, we can expect the theory to have more degrees of freedom. An interesting case is when the phase space is truly infinite-dimensional. 

Let us focus on a particular example of $a(z)$ with infinite number of zeros,
\begin{equation}\label{eq:sinxx}
a(z)=\frac{\sin{(z)}}{z}=1+\sum_{j=1}^\infty\frac{(-1)^{j}z^{2j}}{(2j{+}1)!}=\prod_{k=1}^\infty\Big(1-\frac{z^2}{\pi^2 k^2}\Big)\;.
\end{equation}
The free part of the Lagrangian \eqref{eq:lagrangian} in the ${(1{+}1)}$-dimensional language is then
\begin{equation}
	\hat{\mathscr{L}}[Q](t,s) = -\frac12 Q(t,s) \sin(\pp_s^2)Q(t,s)\;.
\end{equation}
Note that this Lagrangian still contains just the kinetic terms and no potential term because the lowest term in expansion of $\sin(\pp_s^2)$ is $\pp_s^2$, cf. Eq.~\eqref{eq:sinxx}. Recalling the expressions Eqs.~\eqref{eq:psifree} and \eqref{eq:phifree}, which are valid for arbitrary analytic function $a(z)$, we arrive at the Euler--Lagrange constraint,
\begin{equation}\label{eq:ELsin}
\Psi(t,s)=-\sin(\pp_s^2)Q(t,s)\approx0\;,
\end{equation}
and the momentum constraint,
\begin{equation}\label{eq:Msin}
\begin{gathered}
\Phi(t,s) =P(t,s){-}\frac12\sum_{j=0}^\infty \frac{(-1)^{j}}{(2j{+}1)!}\!\sum_{l=0}^{2j} Z^{l,2j-l}[Q](t,s)\approx0\;.
\end{gathered}
\end{equation}
The constraints $\Psi(t,s)$ and $\Phi(t,s)$ are of the second class, see Appx.~\ref{ap:seccl}.

We need to find a general solution of the Euler--Lagrange constraint Eq.~\eqref{eq:ELsin}. The function $1/\sin(z^2)$ has a second-order pole at ${z=0}$ and infinite number of simple poles at ${z=\pm\sqrt{\pi k}}$ and ${z=\pm i\sqrt{\pi k}}$, ${k\in\mathbb{N}}$. Suppressing the $t$-dependence, the analyticity of $\sin(z^2)\mathcal{L}[Q](z)$ implies
\begin{equation}
\begin{aligned}
\mathcal{L}[Q](z) &=\frac{A_0}{z}+\frac{B_0}{z^2}+\sum_{k=1}^\infty\bigg[\frac{B_k}{z{-}\sqrt{\pi k}}+\frac{B_{k}}{z{+}\sqrt{\pi k}}
\\
&\feq+\frac{A_{-k}}{z{-}i\sqrt{\pi k}}+\frac{A_{-k}}{z{+}i\sqrt{\pi k}}\bigg]\;,
\end{aligned}
\end{equation}
for some arbitrary constants $A_j$ and $B_j$, ${j\in\mathbb{Z}}$. After performing the inverse Laplace transform, calculating the integrals by means of the Cauchy integral theorem, and restoring the $t$-dependence, we arrive at
\begin{equation}
\begin{aligned}
Q(t,s)&= \alpha_0(t) {+} \beta_0(t)s 
\\
&\feq+\sum_{k=1}^\infty\big[\tfrac{\alpha_k(t){+}\beta_k(t)}{2}e^{\sqrt{\pi k}s}{+}\tfrac{\alpha_k(t){-}\beta_k(t)}{2}e^{-\sqrt{\pi k}s}
\\
&\feq+\alpha_{-k}(t)\cos{(\sqrt{\pi k}s)}+\beta_{-k}(t)\sin{(\sqrt{\pi k}s)}\big]\;,
\end{aligned}
\end{equation}
where $\alpha_j(t)$ and $\beta_j(t)$, ${j\in\mathbb{Z}}$, are arbitrary functions. The coefficients were chosen in order to get simple expressions for derivatives of ${Q(t,s)}$ at ${s=0}$,
\begin{equation}
\begin{aligned}
Q^{(2j)}_{s{=}0} &=\alpha_0 \de_0^j+\sum_{k=1}^\infty\big[\alpha_k+(-1)^j\alpha_{-k}\big](\pi k)^{j}\;,
\\
Q^{(2j{+}1)}_{s{=}0} &=\beta_0 \de_0^j+\sum_{k=1}^\infty\big[\beta_k+(-1)^j\beta_{-k}\big](\pi k)^{j+\frac12}\;.
\end{aligned}
\end{equation}

Instead of solving the momentum constraint Eq.~\eqref{eq:Msin} for $P(t,s)$, we can insert these expressions directly in the general formulas Eq.~\eqref{eq:HOalter}. After some algebra, the resulting Hamiltonian and the symplectic 2-form on the reduced phase space get the simple form
\begin{equation}\label{eq:hamomsin}
\begin{aligned}
H_\mathrm{red} &=\frac12 \beta_0^2+ \sum_{k=1}^\infty (-1)^k\pi k\big(-\alpha_k^2+\beta_k^2+\alpha_{-k}^2+\beta_{-k}^2\big) \;,
\\
\Omega_\mathrm{red} &=\dd \alpha_0 {\bwedge}\dd \beta_0
+\!\!\sum_{k=1}^\infty (-1)^k \big( \dd \alpha_k {\bwedge}\dd \beta_k +\dd \alpha_{-k} {\bwedge}\dd \beta_{-k}\big)\;.
\end{aligned}
\end{equation}
From the symplectic 2-form, we can conclude that the reduced phase space, parametrized by variables $\alpha_j$ and $\beta_j$, ${j\in\mathbb{Z}}$, is truly infinite-dimensional. Therefore, the theory has infinite number of dynamical degrees of freedom.


\section{Summary}\label{sc:sum}

In this paper, we studied the non-local scalar field theory Eq.~\eqref{eq:actionphi}, which is obtained by metric perturbation of the infinite derivative gravity action Eq.~\eqref{eq:SIDG}. The equations of motion are assumed to have the same scaling symmetry as Einstein's equations. We focused on spatially homogeneous fields.

First, we analyzed the reduced phase space of the free theory with $a(z)$ being an entire function with no-zeros in the complex plane. We confirmed an expected result that such a system is dynamically equivalent to a local theory with one degree of freedom. Then, we investigated the theory with interactions by expanding the solution in the coupling constant. By iteratively solving the constraints, we computed the perturbative reduced Hamiltonian Eq.~\eqref{eq:hamom}. Finally, we discussed an illustrative example where $a(z)$ has infinite number of zeros, ${a(z)=\sin(z)/z}$. This choice leads to an infinite-dimensional reduced phase space, see Eq.~\eqref{eq:hamomsin}.

There are many open questions which are hard and require further investigation. For example, we would like to include the spatially non-homogeneous fields. Also, it would be very interesting to explore the covariant formulation of Hamiltonian construction for non-local theories. This step is necessary for the study of the full action of the infinite derivative gravity.


\section*{Acknowledgements}
Authors would like to thank Joaquim Gomis, Pavel Krtou\v{s}, Ashoke Sen, Tirthabir Biswas, and Alexey Koshelev for valuable discussions.

I.K. and A.M. are supported by Netherlands Organization for Scientific Research (NWO) grant no. 680-91-119.


\appendix

\section{Second-class character of constraints} \label{ap:seccl}
In the following, we demonstrate that the set of Euler--Lagrange and momentum constraints (with the continuous index $s$)
\begin{equation}\label{eq:ELMC}
\begin{aligned}
\Psi(t,s) &=-a(\pp_s^2)Q''(t,s)\approx0\;,
\\
\Phi(t,s) &=P(t,s)-\frac12\sum_{k,j=0}^\infty a_{k{+}j}Z^{k,j}[Q](t,s)\approx0\;,
\end{aligned}
\end{equation}
are of the second class for an arbitrary analytic function $a(z)$ with ${a(0)=1}$. For this purpose, we have to compute the matrix
\begin{equation}
C(s,\tilde{s})=\begin{bmatrix}
\pois{\Psi(s)}{\Psi(\tilde{s})}\! & \!\pois{\Psi(s)}{\Phi(\tilde{s})}
\\
\pois{\Phi(s)}{\Psi(\tilde{s})}\! & \!\pois{\Phi(s)}{\Phi(\tilde{s})}
\end{bmatrix}
\end{equation}
and determine its rank.

In order to simplify our calculations, we replace the phase-space quantities by their components in the Taylor basis
\begin{equation}
\begin{aligned}
e^k(s) &=(-1)^k\delta^{(k)}(s)\;,
\\
e_k(s) &=\frac{s^k}{k!}\;.
\end{aligned}
\end{equation}
Note that $e^k(s)$ and $e_k(s)$ satisfy the orthonormality relations
\begin{equation}
\begin{aligned}
\int_\mathbb{R}\!\! ds\,e^k(s)e_l(s) &=\de_l^k\;,
\\\
\sum_{k=0}^\infty e^k(s)e_k(\tilde{s}) &=\de(s-\tilde{s})\;.
\end{aligned}
\end{equation}
The Taylor-basis components of the canonical variables are given by the expansions
\begin{equation}\label{eq:compTayl1}
Q(t,s)=\sum_{k=0}^\infty q^k(t)e_k(s)\;, 
\quad
P(t,s)=\sum_{k=0}^\infty p_k(t)e^k(s)\;.
\end{equation}
The new quantities $q^k$ and $p_k$ play a role of the canonical coordinates on the phase space.

The components Euler--Lagrange and momentum constraints read 
\begin{equation}\label{eq:compTayl2}
\Psi(t,s)=\sum_{k=0}^\infty \Psi^k(t)e_k(s)\;,
\quad
\Phi(t,s)=\sum_{k=0}^\infty \Phi_k(t)e^k(s)\;.
\end{equation}
 
By inserting the expansions Eq.~\eqref{eq:compTayl1} in Eq.~\eqref{eq:omega}, we obtain the expression for the symplectic 2-form,
\begin{equation}
\Omega=\sum_{k=0}^\infty \dd q^k\bwedge \dd p_k\;.
\end{equation}
The inversion of $\Omega$ provides the formula for the Poisson bracket of two phase-space observables ${F=F(q^j,p_k)}$ and ${G=G(q^j,p_k)}$, cf. Eq.~\eqref{eq:pois},
\begin{equation}\label{eq:poiscom}
\pois{F}{G} = \sum_{k=0}^\infty \bigg[\frac{\pp F}{\pp q_k}\frac{\pp G}{\pp p^k}-\frac{\pp F}{\pp p_k}\frac{\pp G}{\pp q^k} \bigg]\;.
\end{equation}
The matrix $C$ can be equivalently written by means of the components form$\Psi^k$ and $\Phi_k$,
\begin{equation}\label{eq:Cmatcomp}
C_{mj} = 
\begin{bmatrix}
\pois{\Psi^m}{\Psi^j}\! & \!\pois{\Psi^m}{\Phi_j}
\\
\pois{\Phi_m}{\Psi^j}\! & \!\pois{\Phi_m}{\Phi_j}
\end{bmatrix}.
\end{equation}

Employing Eqs.~\eqref{eq:compTayl1} and \eqref{eq:compTayl2}, we obtain the components of the constraint Eq.~\eqref{eq:ELMC} in terms of $q^j$ and~$p_k$,
\begin{equation}
\begin{aligned}
\Psi^m &=-\sum_{k=0}^\infty a_k q^{m{+}2k{+}2}\approx 0\;,
\\
\Phi_{2l} &=p_{2l} - \frac12\sum_{k=0}^\infty a_{k{+}l}q^{2k{+}1}\approx 0\;,
\\
\Phi_{2l{+}1} &=p_{2l{+}1} + \frac12\sum_{k=0}^\infty a_{k{+}l}q^{2k}\approx 0\;.
\end{aligned}
\end{equation}
After evaluating all combination of the Poisson brackets using Eq.~\eqref{eq:poiscom}, we get the explicit expression for the matrix~Eq.~\eqref{eq:Cmatcomp},
\begin{widetext}
\begin{equation}\label{eq:Cbigmat}
C =
\begin{bmatrix}
0 & 0 & 0 & 0  & \!\dots\! & 0 & 0 & {-}a_0 & 0 & {-}a_1 & 0 & \!\dots
\\
0 & 0 & 0 & 0  & \!\dots\! & 0 & 0 & 0 & {-}a_0 & 0 & {-}a_1 & \!\dots
\\
0 & 0 & 0 & 0  & \!\dots\! & 0 & 0 & 0 & 0 & {-}a_0 & 0 & \!\dots
\\
0 & 0 & 0 & 0  & \!\dots\! & 0 & 0 & 0 & 0 & 0 & {-}a_0 & \!\dots
\\
\vdots & \vdots & \!\vdots\! & \vdots  & \ddots & \vdots & \vdots & \vdots & \vdots & \vdots & \vdots & \!\ddots
\\
0 & 0 & 0 & 0 & \!\dots\! & 0 & {-}a_0 & 0 & {-}a_1 & 0 & {-}a_2 & \!\dots
\\
0 & 0 & 0 & 0 & \!\dots\! & a_0 & 0 & a_1 & 0 & a_2 & 0 & \!\dots
\\
a_0 & 0 & 0 & 0 & \!\dots\! & 0 & {-}a_1 & 0 & {-}a_2 & 0 & {-}a_3 & \!\dots
\\
0 & a_0 & 0 & 0 & \!\dots\! & a_1 & 0 & a_2 & 0 & a_3 & 0 & \!\dots
\\
a_1 & 0 & a_0 & 0 & \!\dots\! & 0 & {-}a_2 & 0 & {-}a_3 & 0 & {-}a_4 & \!\dots
\\
0 & a_1 & 0 & a_0 & \!\dots\! & a_2  & 0 & a_3 & 0 & a_4 & 0 & \!\dots
\\
\vdots & \vdots & \vdots & \vdots & \!\ddots\!  & \vdots & \vdots & \vdots & \vdots & \vdots & \vdots & \!\ddots
\end{bmatrix}
\sim 
\begin{bmatrix}
0 & 0 & 0 & 0  & \dots & 0 & 0 & {-}1 & 0 & {-}a_1 & 0 & \!\dots
\\
0 & 0 & 0 & 0  & \dots & 0 & 0 & 0 & {-}1 & 0 & {-}a_1 & \!\dots
\\
0 & 0 & 0 & 0  & \dots & 0 & 0 & 0 & 0 & {-}1 & 0 & \!\dots
\\
0 & 0 & 0 & 0  & \dots & 0 & 0 & 0 & 0 & 0 & {-}1 & \!\dots
\\
\vdots & \vdots & \vdots & \vdots  & \ddots & \vdots & \vdots & \vdots & \vdots & \vdots & \vdots & \!\ddots
\\
0 & 0 & 0 & 0 & \dots & 0 & {-}1 & 0 & 0 & 0 & 0 & \!\dots
\\
0 & 0 & 0 & 0 & \dots & 1 & 0 & 0 & 0 & 0 & 0 & \!\dots
\\
1 & 0 & 0 & 0 & \dots & 0 & 0 & 0 & 0 & 0 & 0 & \!\dots
\\
0 & 1 & 0 & 0 & \dots & 0 & 0 & 0 & 0 & 0 & 0 & \!\dots
\\
a_1 & 0 & 1 & 0 & \dots & 0 & 0 & 0 & 0 & 0 & 0 & \!\dots
\\
0 & a_1 & 0 & 1 & \dots & 0 & 0 & 0 & 0 & 0 & 0 & \!\dots
\\
\vdots & \vdots & \vdots & \vdots & \ddots  & \vdots & \vdots & \vdots & \vdots & \vdots & \vdots & \!\ddots
\end{bmatrix}.
\end{equation}
\end{widetext}
Taking into account ${a_0=1\neq0}$, we reduced the matrix by adding appropriate multiples of rows from the upper part of $C$ to its bottom part, followed by adding multiples of the first two rows of the bottom part of $C$ to the rows below them. By applying this similarity transformation we arrive at a matrix that is obviously non-singular. (It could be brought to the identity matrix by further row operations.) Therefore, we have verified that the constraints Eq.~\eqref{eq:ELMC} form a second class set.


%

\end{document}